# Real-time single-molecule imaging of quantum interference

Thomas Juffmann[1], Adriana Milic[1], Michael Müllneritsch[1], Peter Asenbaum[1], Alexander Tsukernik[2],

Jens Tüxen[3], Marcel Mayor[3,4], Ori Cheshnovsky[2,5] and Markus Arndt[1]

[1]*Vienna Center of Quantum Science and Technology, Faculty of Physics, University of Vienna,*

*Boltzmanngasse 5, 1090 Vienna, Austria,* contact: *markus.arndt@univie.ac.at*

[2]*The Center for Nanoscience and Nanotechnology,* Tel Aviv University, 69978 Tel Aviv, Israel

[3]*Department of Chemistry, University of Basel, St. Johannsring 19, 4056 Basel, Switzerland,*

[4]*Karlsruhe Inst. of Technology, Inst. for Nanotechnology, P.O. Box 3640, 76021 Karlsruhe, Germany*

[5]*School of Chemistry, The Raymond and Beverly Sackler faculty of exact sciences, Tel Aviv University, 69978 Tel Aviv, Israel*

The observation of interference patterns in double-slit experiments with massive particles is generally regarded as the ultimate demonstration of the quantum nature of these objects. Such matter-wave interference has been observed for electrons[1], neutrons[2], atoms[3,4] and molecules[5-7] and it differs from classical wave-physics in that it can even be observed when single particles arrive at the detector one by one. The build-up of such patterns in experiments with electrons has been described as the "most beautiful experiment in physics" [8-11]. Here we show how a combination of nanofabrication and nanoimaging methods allows us to record the full two-dimensional build-up of quantum diffraction patterns in real-time for phthalocyanine molecules PcH2 and their tailored derivatives F24PcH2 with a mass of 1298 amu. A laser-controlled micro-evaporation source was used to produce a beam of molecules with the required intensity and coherence and the gratings were machined in 10-nm thick silicon nitride membranes to reduce the effect of van der Waals forces. Wide-field fluorescence microscopy was used to detect the position of each molecule with an accuracy of 10 nm and to reveal the build-up of a deterministic ensemble interference pattern from stochastically arriving and internally hot single molecules.

When Richard Feynman described the double-slit experiment with electrons as "at the heart of quantum physics"[12] he was emphasizing the fundamentally non-classical nature of the superposition principle which allows the quantum wave function associated with a massive object to be widely delocalized, while the object itself is always observed as a well-localized particle. Several recent experiments contributed to a further sharpening of the discussion by demonstrating the stochastic build-up of interferograms[11,13], by implementing double-slit diffraction in the time-domain [14,15], even down to the attosecond level [16], and by identifying a single molecule as the smallest double-slit for electron interference [17,18] that enables fundamental decoherence studies [19]. The extension of far-field diffraction[20] to large molecules requires a sufficiently intense and coherent beam of slow and neutral molecules, a nanosized diffraction grating and a detector with both a spatial accuracy of a few nanometers and a molecule specific detection efficiency of close to 100 %. Our present experiment solves all these tasks simultaneously, using advanced micro-preparation, nanodiffraction and nanoimaging technologies. It thus exposes the quantum wave-particle duality in a particularly clear way and opens the way to new studies with ever larger molecules in an ongoing exploration of the quantum-classical borderline.

Our setup is shown in Figure 1. It is divided into three parts: the beam preparation, coherent manipulation and detection. We need to prepare the molecules such that each of them interferes with itself and that all of them lead to similar interference patterns on the screen. Since the transverse and longitudinal coherence functions are determined by the Fourier transforms of the source spatial extension and velocity distribution [21], we require a good collimation and velocity selection.
As long as we can approximate the molecular wave functions as plane waves – i.e. under 'far-field' conditions – the angle $\theta_n$ to the n-th diffraction order is given by the textbook equation
$\sin \theta_n = n \cdot \lambda / d$, with *d* the grating period, $\lambda$ = h/mv the de Broglie wavelength, Planck's constant *h* and

the molecular momentum $p = m \cdot v$. Massive particles therefore need to be slow to achieve sizeable diffraction angles. Although deceleration techniques have been advanced for molecules even as complex as benzonitrile[22], effusive beams (Figure 1b) are still well suited for preparing slow beams of particles hundred times more massive than that [23,24].

For thermo-labile organic molecules which may decompose when heated to their evaporation temperature we have implemented a laser micro-source (Figure 1a) which reduces the heat-load to a minimum. A blue diode laser is focused onto a thin layer of molecules deposited on the inside of the entrance vacuum window $W_1$ which can be moved by a motorized translation stage. Although high temperatures can be reached locally, this affects only the particles within the focus area. In comparison to a Knudsen cell the heat-load to the sample is thus reduced by 2-3 orders of magnitude, to several 10 mW. Spectral coherence is achieved by sorting the arriving molecules according to their longitudinal velocity and their respective free-fall height in the earth's gravitational field[25].

The collimation slit S defines the spatial coherence of the molecular beam. The slit and the grating width further downstream narrow the beam divergence to smaller than the diffraction angle. The grating is machined into a thin $SiN_x$ membrane and has a period of $d = 100$ nm. In order to minimize the dispersive van der Waals (vdW) interaction between the molecules and the grating wall we reduced the grating thickness from 160 nm in earlier diffraction experiments[5,20] to as little as 10 nm in our present setup. This is important for the manipulation of complex molecules, which may exhibit high polarizabilities, permanent and even thermally induced electric dipole moments [26,27]. Each individually diffracted molecule finally arrives at the 170 µm thin quartz plate $W_2$ that seals the detector vacuum chamber against the ambient air. The gradual emergence of the quantum interference pattern is then observed in wide-field fluorescence microscopy of $W_2$.

Imaging of single molecules in the condensed phase started about two decades ago [28] and various methods for sub-wavelength optical imaging have been developed ever since [29]. Here, we utilize a scheme similar to SHRIMP [30], i.e. single-molecule high resolution imaging with photo-bleaching. Even if the point-spread function of an optical emitter is bound to Abbé's diffraction limit, it is still possible to determine its barycenter with nanometer accuracy, if the signal-to-noise ratio is high enough and as long as the point-spread functions of neighboring molecules do not overlap.

We find that phthalocyanines and their derivatives (PcH2 and F24PcH2 as sketched in Figure 1d and 1e) are stable molecules and efficient dyes even in vacuum. The molecular sample on $W_2$ was illuminated under a shallow angle such that the excitation laser did not enter the imaging optics. The fluorescence was collected by a microscope objective, filtered and imaged onto the single-photon sensitive electron multiplying CCD camera.

Figure 2 shows a typical fluorescence image of surface-deposited phthalocyanines, PcH2. We detect around $10^5$ fluorescence photons per molecule before abrupt bleaching or desorption is observed from one frame to the next – in support of the claim that we monitor single molecules and not aggregates. By fitting a 2D Gaussian to each molecular image we can determine its position with 10 nm accuracy. This would even fulfill the detector requirements of matter-wave near-field interferometry[31].

The high detection efficiency exceeds that of electron-impact quadrupole mass spectrometry by more than a factor $10^4$. This large gain allows us now for the first time to optically visualize the real-time build-up of a 2D quantum interferogram from stochastically arriving single molecules, as shown in Figure 3. This series was recorded with an effusive source (Figure 1b) heated to 750 K. A typical velocity of 150 m/s then corresponds to a de Broglie wavelength of $\lambda_{dB}$ =5.2 pm. The actual velocity distribution is reconstructed from the molecular height distribution on the detection screen and turns out to be slightly faster and narrower than thermal. The pictures represent a balance of continuous accumulation of molecules and intermittent bleaching by the imaging laser (3s per frame for Figure 3a-d). Figure 3 shows the influence of the van der Waals force quite clearly. The high fringe visibility up to the fourth interference order can only be explained by an effective slit narrowing[32] by about a factor of two due to the molecule-wall interaction, even for gratings as thin as 100 atomic monolayers. The relative importance of the molecule-wall interaction is further elucidated in the methods section and in the online supplementary information file as Fig. S2.

The high sensitivity of fluorescence detection now allows us to extend far-field diffraction to more complex molecules, as well. In Figure 4 we compare specifically the interferogram of the fluoroalkylated phthalocyanines F24PcH2 with those of PcH2, both starting from the new laser microevaporation source (Figure 1a) which allows us to record the interferograms with a material consumption 100 times smaller than using the Knudsen cell.

To account for the higher polarizability of F24PcH2 this experiment was performed with somewhat wider slits (75 nm) than used for Figure 3 (s=50 nm). Again, we see clear quantum interference. We retrieve 1D-projections from the 2D diffraction patterns by vertically integrating over a part of the velocity distribution (Fig. 4c and 4d). The solid lines in these diagrams represent the textbook-like diffraction of plane waves at a grating. Additionally, they include an incoherent average over the known source extension as well as over the detected velocity range. We find agreement between the numerical model and our experiment if we fit a van der Waals constant of $C_3$= 16 meV·nm$^3$ for PcH2 and $C_3$= 98 meV·nm$^3$ for F24PcH2 in interaction with the SiN$_x$ walls. The uncertainty in this fit is still about 50%. Future precision measurements of $C_3$ will become possible with a more accurate determination of the open slit width across the entire grating, a better velocity selection and by systematically varying the grating thickness. In order to obtain the excellent fit in Figure 4d it was necessary to convolute the calculated interference pattern with a Gaussian with a standard deviation of 3 µm. This smearing may be attributed either to surface diffusion or to a low contribution of fragmented molecules within our molecular beam. Diffusion would actually be consistent with the design specifications of this molecule: the fluorination is made to reduce the binding to its surroundings, to facilitate the molecular evaporation. Note that in contrast to Figure 3 the patterns of Figure 4 show high contrast only to the first diffraction order. This is related to the larger grating slit width used in this experiment.

In summary, we have shown how a combination of devices produced by nanotechnology permits us for the first time to optically visualize a real-time movie of the wave-particle duality in a textbook-like experiment for large molecules. Compared to our earlier molecular far-field experiments[6], we have improved the source economy by a factor of 1000, reduced the grating thickness and the corresponding van der Waals phase shift by a factor of 16 and increased the detection efficiency to the level of single molecules.

Fluorescence imaging with nanometer accuracy is orders of magnitude more sensitive than the ionization methods used in the past and it is easily scalable: many natural and functionalized organic molecules and quantum-dots are susceptible to this method. In comparison to scanning tunneling microscopy which was previously used for single molecule interference imaging [33] the recording speed can be 1000 times faster for an imaging area which is $10^5$ times larger.

The feasibility of making nanometer thin absorptive gratings promises to push far-field diffraction experiments to new mass limits, in the future[34]. While the van der Waals force is still visible for a membrane as thin as 10 nm, its influence shall be further reduced or completely eliminated in future experiments with masks of double layer graphene or gratings made of light[35].

Grating diffraction of single molecules is an unambiguous and conspicuous demonstration of the wave-particle duality of quantum physics. It is only explicable in quantum terms, independent of the absolute value of the interference contrast. In contrast to photons and electrons which are irretrievably eliminated in the detection process, fluorescent molecules stay in place and can be observed and analyzed over days to show this paradigm of quantum physics.

**Acknowledgements**

This project was funded by the FWF under contract FWF-Z149-N16 (Wittgenstein) and ESF/FWF EuroCore Program MIME (I146). We thank P. Geyer and P. Haslinger for building the in situ sputter cleaning, S. Deachapunya for his collaboration in testing the vapor pressures of PcH2, S. Nimmrichter for theory support and M. Tomandl for rendering Fig 1. MA thanks W.E. Moerner for helpful discussions on single-molecule fluorescence. The chemical synthesis in Basel was supported by the ESF EuroCore Program MIME (I146-N16), the Swiss National Science Foundation and the NCCR 'Nanoscale Science'.

**Author contributions**

TJ and MA conceived the experiments. TJ, AM and MMu and OC worked on the setup of the experiment. TJ performed the diffraction experiments. JT and MMa designed and synthesized the F24PcH2 molecules. AT and OC fabricated the 10 nm diffraction gratings. PA developed the basis for the microevaporation source. MA and TJ wrote the paper with comments by all authors.

**FIGURES:**

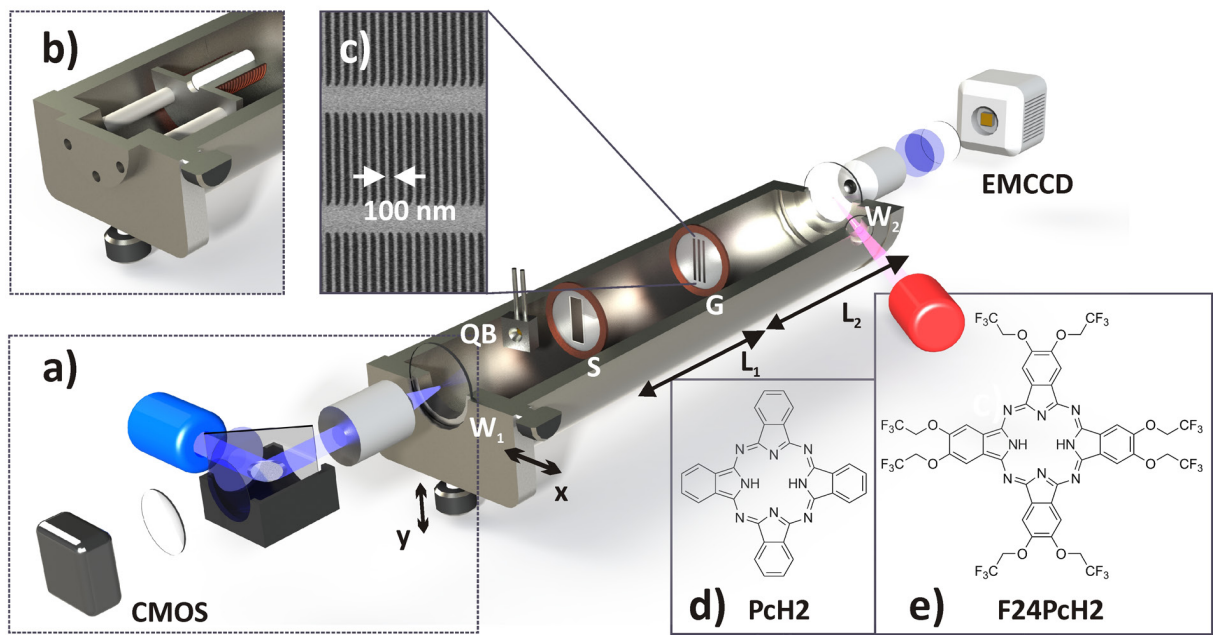

**Figure 1: Setup for the laser-evaporation, diffraction and nanoimaging of complex molecules**

(a) Thermo-labile molecules are ejected by laser micro-evaporation: A blue diode laser (445 nm, 50 mW) is focused onto the window $W_1$ to evaporate the molecules that are coated on its inner surface. A CMOS camera and a quartz balance (QB) monitor the evaporation area and the molecular flux. (b) Stable molecules can be evaporated in a Knudsen cell. The collimation slit S defines the beam coherence. The molecular beam divergence is further narrowed by the width of the diffraction grating G. The grating is nanomachined into a 10 nm thin $SiN_x$ membrane to have a period of d = 100 nm, as shown in the electron micrograph (c). The vacuum system is evacuated to $10^{-8}$ mbar. The molecules on the quartz window $W_2$ are excited by a red diode laser (661 nm). High-resolution optics collects, filters and images the light onto an EMCCD camera. The molecules used in this study are (d) Phthalocyanine PcH2 ($C_{32}H_{18}N_8$, mass m=514 amu, number of atoms N=58) and (e) F24PcH2 ($C_{48}H_{26}F_{24}N_8O_8$, m= 1298 amu, N=114), which doubles the mass, atom number and internal complexity.

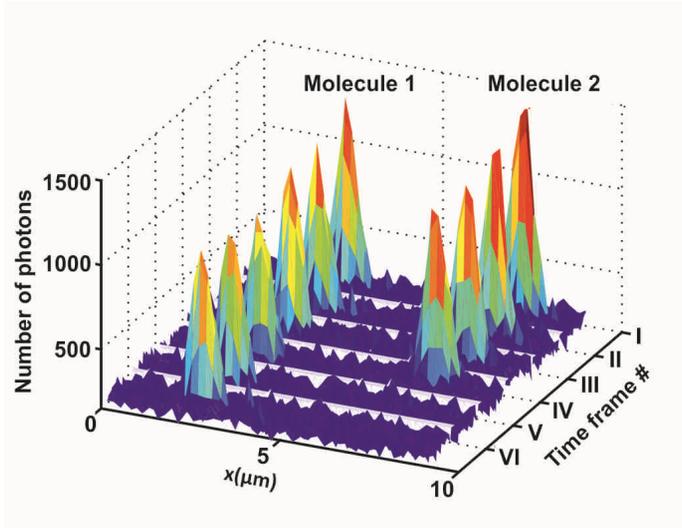

**Figure 2: Single molecule imaging of phthalocyanine PcH2 with sub-wavelength accuracy.**

The figure represents several frames (I to VI from first to last) of a movie recorded with the EMCCD camera. Two molecules are localized on the quartz surface next to each other. After frame IV the molecule 2 either bleaches or desorbs again. Our experiments indicate that bleaching typically occurs after scattering of about $10^5$ photons. We detect each molecule with a signal to noise ratio of about 20, which enables us to determine the barycenter of their point spread function with an accuracy of about 10 nm. Most molecules remain immobilized on the nanoscale and the interference pattern persists even over days.

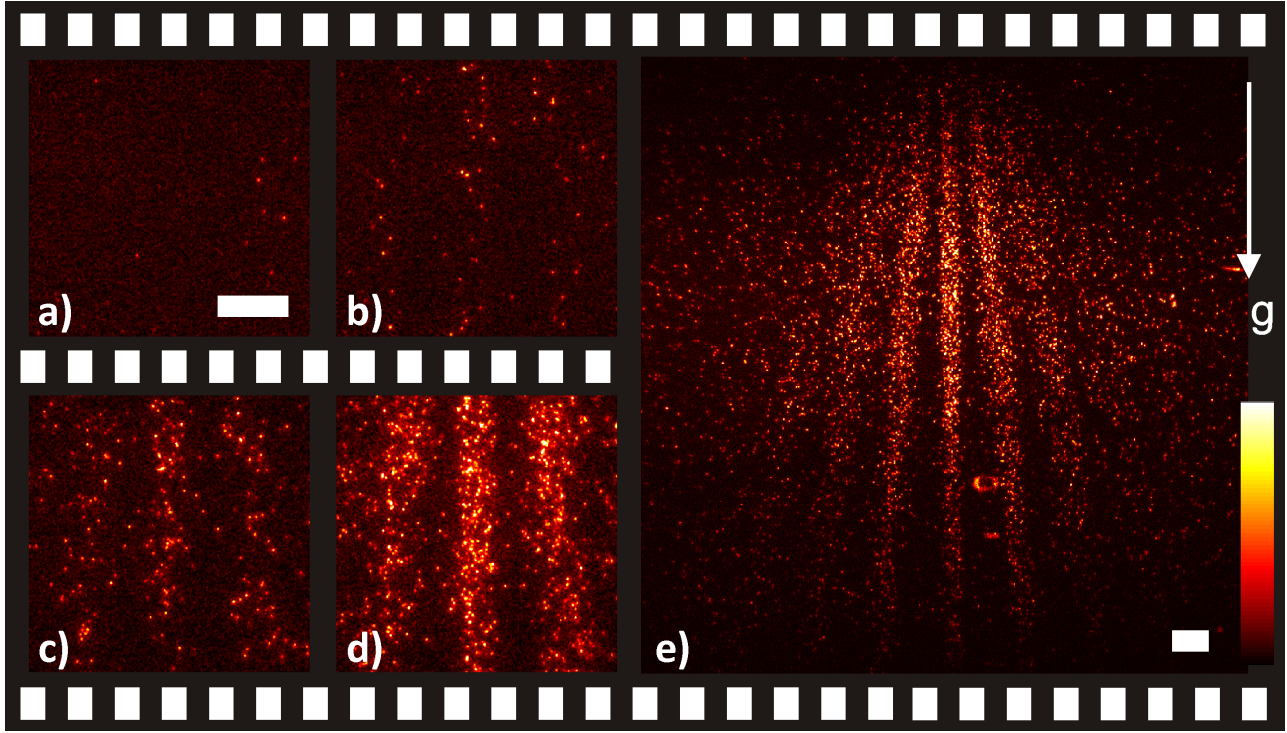

**Figure 3: Selected frames of the quantum molecular movie**

Nanoimaging is the key to visualizing the build-up of a deterministic interference pattern from stochastically arriving single molecules. It is an unambiguous tool for illustrating the quantum wave-particle duality, here of phthalocyanine. The images were recorded a) before deposition and after PcH2 deposition for b) 2 min, c) 20 min, d) 40 min and e) 90 min. The scale bar in Fig. 3e is 20 µm wide. The color bar ranges from 0 to 100 photons for (a-d) and from 0 to 600 photons in e). All details of the imaging parameters are described in the methods section. Two movies are available as supplementary information. Movie 1 contains the frames shown above. Movie 2 is recorded with lower magnification to present a wide-field view as shown in Fig 3e.

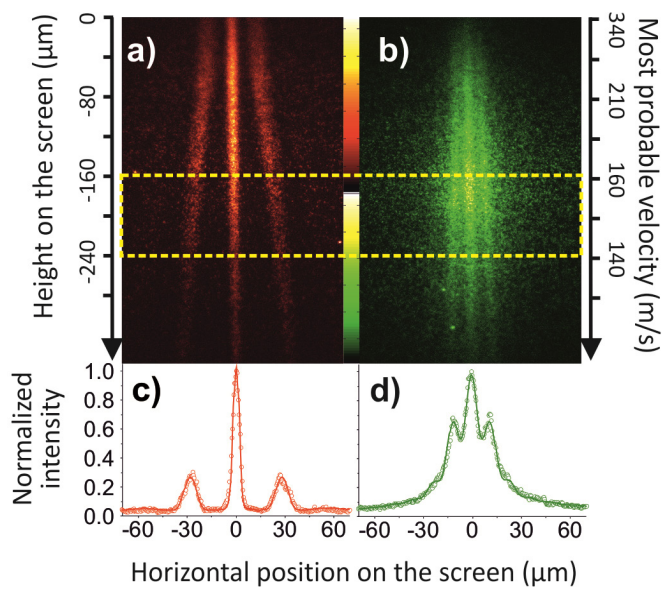

**Figure 4: Comparison of the interference patterns of PcH2 and F24PcH2**

Quantum diffraction fans out the molecular beam in the horizontal direction. Gravity sorts the molecular beam by velocity from fast at the top to slow at the bottom.

*Top:* Fluorescence images of the quantum interferograms of a) PcH2 and b) F24PcH2. From these images we deduce both the mass and the velocity of the molecules. In addition one can deposit such interference patterns next to each other, such that one of them can serve as a reference for the other.

*Bottom:* Integration over $\Delta h$ = 80 µm, corresponding to $\Delta v/v$ = 0.27, results in a 1D diffraction curve. The PcH2 quantum interference pattern (c) shows a fringe contrast close to 100 % and is well reproduced by a numerical model that takes into account the finite source size and van der Waals forces. The F24PcH2 diffraction curve (d) is well reproduced by allowing for an additional 3 µm small spread on the detector. The modeling is described in the Methods section and accompanied by details in the Supplementary Information file. The false colors serve to distinguish the two molecular species. Their true fluorescence of both starts in the range beyond 700 nm. All imaging settings are specified in the Table ST1 of the supplementary methods file.

**Methods**

- *Molecular synthesis:*

  The three step synthesis of F24PcH2 comprised the formation of a fluorinated phthalonitrile followed by the assembly of a fluorous zinc phthalocyanine derivative and a final demetallation step [36] (see Supplementary Methods for synthetic protocols and analytical data).

- *Molecular sample preparation:*

  A solution (PcH2)/suspension (F24PcH2) of molecules in acetone was smeared onto the BK7 entrance window to form a thin layer. Inhomogeneities in the sample thickness may occur but do not influence the final diffraction pattern. We observed some liquefaction during the microevaporation which also homogenized the sample.

- *Fabrication of the nanogratings:*

  The $SiN_x$ gratings were produced by the focused ion beam (FIB) milling into a 10 nm thin membrane purchased from TEMWINDOWS.com. The FIB milling was done in the ionLiNE system (from RAITH GmbH) using Gallium ions at E= 35 keV and with currents ranging from 1 pA to 7 pA. The gratings have a period of 100 nm and an opening width of 50 nm (Figure 4a) and 75 nm (Figure 4b).

- *Surface cleaning by In-situ plasma cleaning* of the quartz window.

  Outside: air at atmospheric pressure. Inside: nitrogen at 1 mbar. Discharge: AC-voltage of 1.5 kV and 10 kHz, 0.5 mm electrode in 0.5 mm distance from the window. The grounded vacuum chamber served as the counter-electrode.

- *Equipment and settings of the imaging optics for all experiments:*

  All equipment, software and parameter specifications are summarized in the "online Supplementary information" Table S3.

- *Experimental parameters of Figure 3:*

  For the first 20 min (a-c) the movie frame rate was 0.1 Hz. After that (d) it was decreased to 0.05 Hz to allow for another dynamic equilibrium of bleaching and fresh arrival of molecules (see the movie in the supplement). The final image e) was recorded after 90 min. The slit S was cut into a t=50 nm thin SiN membrane with a window size of w= 1 × 100 µm². The dimensions of the grating G were: t=10 nm, w=5 × 100 µm², period d=100 nm, open slit width s=50 nm. L1=702 mm, L2=564 mm.

- *Experimental parameters for Figure 4:*

  Collimation slit S: w = 3 µm (defined by a pair of steel razor blades with a 300 nm edge width). Grating G: w=3 × 100 µm², t=10 nm, d = 100 nm, s=75 nm. L1=566 mm, L2=564 mm.

- *Numerical modeling of the diffraction images:*

  We fit our data using diffraction integrals in the paraxial (Eikonal) approximation. In the last step, this involves an incoherent sum over all coherent diffraction patterns that are associated with molecules starting from different source points – causing limited spatial coherence – and with different velocities which cause limited spectral coherence. To identify the contributing velocities we fit the molecular distribution on the screen to a Maxwell-Boltzmann velocity and take into account all vertical constraints in our setup. We can thus assign the forward velocity of a molecule according to its vertical position on the screen. The van der Waals interaction between the polarizable molecule and the dielectric grating wall is taken into account in the phase of the grating transmission function. The phase term $\psi(\phi) = \exp(i/\hbar * t/v * C_3 * [1/\Delta x^3 + 1/(s-\Delta x)^3])$ is multiplied to the binary transmission function given by the grating's period and opening fraction. Here, $C_3$ denotes the van der Waals constant, which we determine from a numerical fit of the expected diffraction curve to the observed interference pattern. The distance Δx of a particular molecule to its nearest grating bar as well as its longitudinal velocity are important for the effective momentum kick during the passage through the grating. We assume Δx to be constant for each molecule during the transit time through the grating. We neglect all fringe effects, such as the attraction outside of the grating slit. In order to illustrate the high significance of surface-wall interactions even for a grating as thin as 10 nm and a molecular transit time of only 100 ps, we compare our experimental data to different theoretical assumptions in the "Supplementary information" file, Figure S2.

**References**

1.  Jönsson, C. Elektroneninterferenzen an mehreren künstlich hergestellten Feinspalten. *Z. Phys.* **161**, 454-474 (1961).
2.  Zeilinger, A., Gähler, R., Shull, C. G., Treimer, W. & Mampe, W. Single- and Double-Slit Diffraction of Neutrons. *Rev. Mod. Phys.* **60**, 1067-1073 (1988).
3.  Keith, D. W., Schattenburg, M. L., Smith, H. I. & Pritchard, D. E. Diffraction of Atoms by a Transmission Grating. *Phys. Rev. Lett.* **61**, 1580-1583 (1988).
4.  Carnal, O. & Mlynek, J. Young's Double-Slit Experiment with Atoms: A Simple Atom Interferometer. *Phys. Rev. Lett.* **66**, 2689-2692 (1991).
5.  Schöllkopf, W. & Toennies, J. P. Nondestructive Mass Selection of Small Van der Waals Clusters. *Science* **266**, 1345-1348 (1994).
6.  Arndt, M. *et al.* Wave-particle duality of C-60 molecules. *Nature* **401**, 680-682 (1999).
7.  Zhao, B. S., Meijer, G. & Schollkopf, W. Quantum reflection of He2 several nanometers above a grating surface. *Science* **331**, 892-894, doi:10.1126/science.1200911 (2011).
8.  Crease, R. P. The most beautiful experiment in physics. *Phys. World* (2002).
9.  *The double-slit experiment Phys. World 15, 13 (September 2002); physicsworld.com/cws/article/print/9745*
10. Merli, P., Missiroli, G. & Pozzi, G. On the statistical aspect of electron interference phenomena. *American Journal of Physics* **44**, 306 (1976).
11. Tonomura, A,., Endo, J,., Matsuda, T,., Kawasaki, T,. & Ezawa, H,. Demonstration of Single-Electron Buildup of an Interference Pattern. *Am. J. Phys.* **57**, 117-120 (1989).
12. Feynman, R., Leighton, R. B. & Sands, M. L. *The Feynman Lectures on Physics, Vol III, Quantum Mechanics*. (Addison Wesley, Reading (Mass), 1965).
13. Juffmann, T. *et al.* Wave and Particle in Molecular Interference Lithography. *Phys. Rev. Lett.* **103**, 263601 (2009).
14. Szriftgiser, P., Guéry-Odelin, D., Arndt, M. & Dalibard, J. Atomic Wave Diffraction and Interference Using Temporal Slits. *Phys. Rev. Lett.* **77**, 4-7 (1996).
15. Garcia, N., Saveliev, I. G. & Sharonov, M. Time-resolved diffraction and interference: Young's interference with photons of different energy as revealed by time resolution. *Philos. Transact. A. Math. Phys. Eng. Sci.* **360**, 1039-1059, doi: (2002).
16. Lindner, F. *et al.* Attosecond double-slit experiment. *Phys. Rev. Lett.* **95**, 40401 (2005).
17. Akoury, D. *et al.* The simplest double slit: Interference and entanglement in double photoionization of H-2. *Science* **318**, 949-952 (2007).
18. Canton, S. E. *et al.* Direct observation of Young's double-slit interferences in vibrationally resolved photoionization of diatomic molecules. *Proceedings of the National Academy of Sciences* **108**, 7302-7306, doi: (2011).
19. Zimmermann, B. *et al.* Localization and loss of coherence in molecular double-slit experiments. *Nature Physics* **4**, 649-655 (2008).
20. Nairz, O., Arndt, M. & Zeilinger, A. Quantum Interference Experiments with Large Molecules. *Am. J. Phys.* **71**, 319 (2003).
21. Born, M. & Wolf, E. *Principles of Optics*. (Pergamon Press, 1993).
22. Wohlfart, K. *et al.* Alternating-gradient focusing and deceleration of large molecules. *Phys. Rev. A* **77**, 031404R (2008).
23. Deachapunya, S. *et al.* Slow beams of massive molecules. *The European Physical Journal D* **46**, 307-313 (2007).

24  Gerlich, S. *et al.* Quantum interference of large organic molecules. *Nature Communs.* **2**, 263, doi:10.1038/ncomms1263 (2011).
25  Nairz, O., Arndt, M. & Zeilinger, A. Experimental Challenges in Fullerene Interferometry. *J. Mod. Opt.* **47**, 2811-2821 (2000).
26  Compagnon, I., Antoine, R., Rayane, D., Broyer, M. & Dugourd, P. Vibration Induced Electric Dipole in aWeakly Bound Molecular Complex. *Phys. Rev. Lett.* **89** 253001 (2002).
27  Gring, M. *et al.* Influence of conformational molecular dynamics on matter wave interferometry. *Phys.Rev. A* **81**, 031604 (2010).
28  Moerner, W. E. & Kador, L. Optical detection and spectroscopy of single molecules in a solid. *Phys. Rev. Lett.* **62** 2535 - 2538 (1989).
29  Hell, S. W. & Wichmann, J. Breaking the diffraction resolution limit by stimulated emission: stimulated-emission-depletion fluorescence microscopy. *Optics Letters* **19**, 780–782 (1994).
30  Gordon, M. P., Ha, T. & Selvin, P. R. Single-molecule high-resolution imaging with photobleaching. *PNAS* **101**, 6462 - 6465 (2004).
31  Gerlich, S. *et al.* A Kapitza-Dirac-Talbot-Lau interferometer for highly polarizable molecules. *Nature Phys.* **3**, 711 - 715 (2007).
32  Grisenti, R. E., Schöllkopf, W., Toennies, J. P., Hegerfeldt, G. C. & Köhler, T. Determination of Atom-Surface Van der Waals Potentials from Transmission-Grating Diffraction Intensities. *Phys. Rev. Lett.* **83**, 1755 (1999).
33  Juffmann, T. *et al.* Wave and particle in molecular interference lithography. *Phys. Rev. Lett.* **103**, 263601 (2009).
34  Juffmann, T., Nimmrichter, S., Arndt, M., Gleiter, H. & Hornberger, K. New prospects for de Broglie interferometry *Found. Phys.* (2010).
35  Nairz, O., Brezger, B., Arndt, M. & Zeilinger, A. Diffraction of Complex Molecules by Structures Made of Light. *Phys. Rev. Lett.* **87**, 160401 (2001).
36  Alzeer, J., Roth, P. J. C. & Luedtke, N. W. An efficient two-step synthesis of metal-free phthalocyanines using a Zn(ii) template. *Chem. Comm.*, 1970-1971 (2009).